\documentclass[10 pt, conference]{IEEEtran}
\IEEEoverridecommandlockouts                       

\usepackage{url}
\usepackage[utf8]{inputenc}
\usepackage[T1]{fontenc}
\usepackage{amsmath,amssymb,amsfonts}
\usepackage{graphicx}
\usepackage{subfigure}
\usepackage{textcomp}
\usepackage{xcolor}
\usepackage{verbatim}
\usepackage{makecell}
\usepackage{booktabs}
\usepackage{cite}
\usepackage{caption}
\usepackage{extarrows}
\usepackage{bm}
\usepackage{extarrows}
\usepackage{lettrine}
\usepackage[implicit=false]{hyperref}
\usepackage{amsthm}
\usepackage{algorithm,algorithmic}

\begin{document}
\title{\huge
Blind IRS Beamforming Using Only Received Signal Power Measurements
\vspace{-0.5em}} 

\vspace{0.5em}
\author{
\IEEEauthorblockN{Tao Wang\IEEEauthorrefmark{1}, Xiaohui Zhang\IEEEauthorrefmark{1}, Pengfei Lv\IEEEauthorrefmark{1}, Hehe Ban\IEEEauthorrefmark{1}, Yiwei Guo\IEEEauthorrefmark{1}, and Ming Yi\IEEEauthorrefmark{2}
}
\IEEEauthorblockA{\IEEEauthorrefmark{1}Songshan Laboratory, Zhengzhou, China}
\IEEEauthorblockA{\IEEEauthorrefmark{2}Information Engineering University, PLA Cyberspace Force, Zhengzhou, China}
\IEEEauthorblockA{Emails:\{taowang, zhangxiaohui\}@songshanlab.com, lpf19950818@163.com, \{1720826586, 3830875\}@qq.com, \\ and acco666666@sina.com.cn}
\vspace{-0.5em}
\thanks{This work was supported in part by the Mobile Information Networks National Science and Technology Major Project under Grant 2025ZD1303100, and the Henan Provincial Major Science and Technology Project under Grant 241110210300.}

\vspace{-2em}
}
\maketitle
\thispagestyle{empty}
\pagestyle{empty}

\begin{abstract}
This paper proposes a novel blind beamforming strategy for signal enhancement aided by an intelligent reflecting surface (IRS), using only samples of the received signal power. Unlike existing conditional sampling mean (CSM) based approaches, which rely solely on comparing the CSM values, the proposed algorithm fully exploits the information in the sampled data by constructing a least squares problem. This enables the estimation of the phase difference between the IRS-reflected channel and the direct channel, ultimately yielding the optimal IRS phase configuration. The simulation results verify that, by fully leveraging the collected data, the proposed scheme outperforms the CSM-based method, especially when the number of IRS phase quantization levels is larger than 2, while incurring the same level of computational complexity.
\end{abstract}

\begin{IEEEkeywords}
Signal enhancement, intelligent reflecting surface (IRS), blind beamforming, received signal power.
\end{IEEEkeywords}


\section{Introduction}
Intelligent Reflecting Surfaces (IRSs) are widely regarded as a promising technology for next-generation wireless communication systems \cite{BJORNSON-MassiveMIMO-next-2019, you2024generationadvancedtransceivertechnologies,You-IRS-tutorial}. By dynamically tuning the phase shifts of passive elements, IRSs can controllably reshape wireless propagation environments. This capability enables signal enhancement and interference suppression without the need for expensive radio frequency (RF) chains \cite{Power-Scaling-Laws-and-Near-Field-Behaviors, IRS_Anti_Jamming_survey_2025}. 
The performance of IRS critically depends on the precise configuration of phase shifts, i.e., passive beamforming. Conventional methods follow a two-stage paradigm—\emph{channel estimation first, beamforming optimization later}. However, acquiring accurate Channel State Information (CSI) in IRS-aided systems faces major practical challenges \cite{CSM_2023}. First, the weak reflection through a single IRS element is easily overwhelmed by the much stronger direct link and noise, making accurate estimation difficult \cite{CE_single_element}. Second, estimating IRS-related channels usually requires in-phase/quadrature (I/Q) signal components, which are not supported by current network protocols such as 5G \cite{CSM_2023}. Third, channel estimation often involves computationally costly operations like matrix inversion or compressed sensing \cite{IRS_CE_CS}, increasing both complexity and pilot overhead.

These constraints have spurred interest in CSI-free, or blind, beamforming. One simple strategy is random beamforming, which requires no instantaneous CSI \cite{IRS_rotation}. Another line of work relies on beam training, which is mainly suitable for mmWave/THz bands with sharp beams \cite{IRS_BT_tao_2025, multiple_IRS_tao_2025}. 
Departing from these, RFocus \cite{RFocus} leverages statistical features of the received power to configure the IRS. Inspired by this, a blind beamforming algorithm based on the Conditional Sample Mean (CSM) of received power was introduced in \cite{CSM_2023} and later extended to multi-user and multi-antenna scenarios \cite{CSM_multi_UE, CSM_MIMO}. The CSM algorithm selects, for each IRS element, the discrete phase shift that maximizes the corresponding conditional sample mean of received power. Theoretically, with sufficient samples, the CSM configuration converges with high probability to the solution of the perfect-CSI-based Closest Point Projection (CPP) algorithm \cite{CSM_2023}.

Nevertheless, the CSM algorithm relies on directly comparing the magnitudes of conditional sample means across different phase shifts. Consequently, it does not fully exploit the information contained in the sampled power measurements, which may limit performance gains, especially when sampling overhead is constrained or the number of phase quantization levels is large.
To overcome this limitation, we propose a novel blind IRS beamforming strategy. Unlike CSM, the proposed scheme makes full use of the sample data. The core idea is to formulate a least-squares problem that directly estimates the relative phase difference between the background (direct) channel and each IRS-reflected channel using only received signal power measurements. Once estimated, the IRS phase shifts are configured to compensate for these differences, enabling coherent combining and thus maximizing the received signal-to-noise ratio (SNR). 
The main contributions of this work are summarized as follows:
\begin{itemize}
    \item We propose a novel blind IRS beamforming algorithm. By formulating a least-squares problem, it fully extracts phase information from the same statistical measurements used by CSM, leading to more accurate phase estimation.
    \item We show that for $K=2$ phase quantization levels, the proposed algorithm is theoretically equivalent to CSM, providing a unified perspective. For $K>2$, however, it achieves superior performance.
    \item Simulations confirm that by fully leveraging the collected data, the proposed scheme significantly outperforms the CSM-based method, especially for larger $K$. This gain is achieved while maintaining the same order of computational complexity, i.e., $\mathcal{O}(TNK)$.
\end{itemize}

The remainder of the paper is organized as follows. Section~\ref{sec:system_model} presents the system model. Section~\ref{sec:benchmark_schemes} reviews relevant existing schemes. Section~\ref{sec:proposed_scheme} details the proposed blind beamforming algorithm. Section~\ref{sec:simulations} provides numerical results, and Section~\ref{sec:conclusion} concludes the paper.

The reproducible code for our numerical simulations is available at: https://gitee.com/sssystaowang/blind-irs-beamforming-for-signal-enhancement.

\section{System Model}
\label{sec:system_model}
\begin{figure}[t]
\centering
\includegraphics[width=0.4\textwidth]{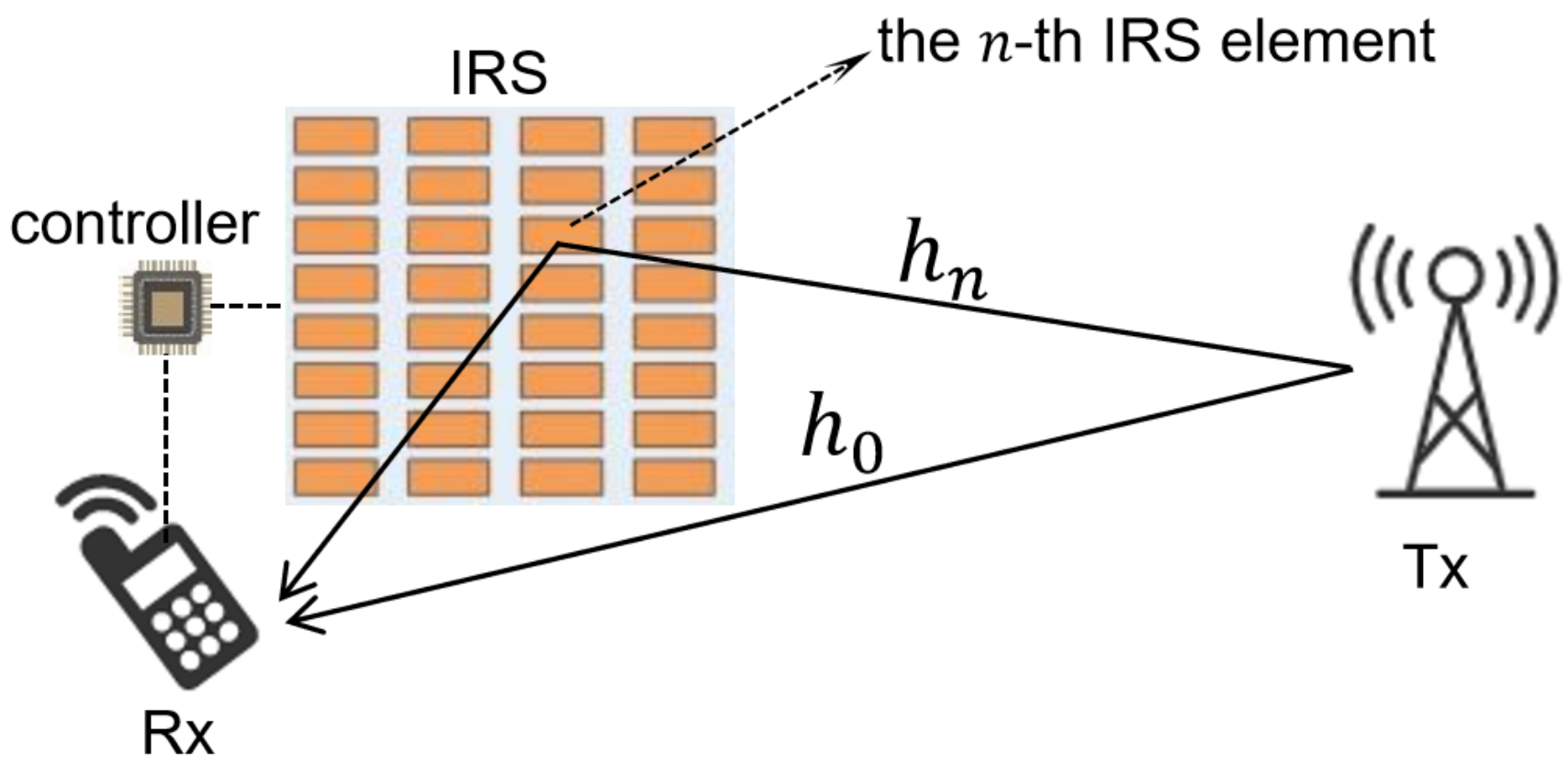}
\caption{An IRS-aided wireless communication system.}
\label{fig-system-model}
\end{figure}

As depicted in Fig.~\ref{fig-system-model}, we consider a wireless communication system where an IRS composed of $N$ passive reflecting elements is deployed to enhance the downlink signal. The IRS is controlled by the receiver (Rx)\footnote{The IRS adjusts its configuration based solely on commands from the Rx, without providing any feedback. This simplified architecture facilitates seamless integration with existing networks.}\footnote{If channel reciprocity holds, the IRS phase configuration obtained will also benefit uplink transmission. Otherwise, uplink enhancement can be achieved by symmetrically deploying the IRS and the proposed algorithm at the transmitter (Tx).}. Let $\mathcal{N} = \{1, 2, \dots, N\}$ index the IRS elements, with $\mathrm{IRS}_n$ denoting the $n$-th element. The received signal at the Rx is expressed as:
\begin{equation}
\label{eq:received_signal}
\begin{split}
  Y &= \left(\beta_{0}e^{j\alpha_0} + \sum_{n=1}^{N} \beta_{n} e^{j\alpha_n}e^{j\theta_{n}}\right) X + Z\\
  &=\left(h_{0} + \sum_{n=1}^{N} h_{n} e^{j \theta_{n}}\right) X + Z,
\end{split}
\end{equation}
where $X$ is the transmitted symbol from the Tx with average power $P$, i.e., $\mathbb{E}\left[|X|^{2}\right] = P$. Here, $h_{0} = \beta_0 e^{j\alpha_0}$ represents the direct (background) channel from the Tx to the Rx, and $h_{n} = \beta_n e^{j\alpha_n}$ denotes the equivalent cascaded channel via $\mathrm{IRS}_n$ (i.e., the Tx-$\mathrm{IRS}_n$-Rx link). The phase shift introduced by $\mathrm{IRS}_n$ is denoted by $\theta_{n} \in [0, 2\pi)$, and $Z\sim \mathcal{CN}\left(0, \sigma^{2}\right)$ is the additive white Gaussian noise. In practice, due to hardware constraints, each phase shift $\theta_{n}$ is selected from a discrete set:
\begin{equation}
\label{eq:phase_set}
\theta_n \in \Phi_{K} = \{\omega, 2\omega, \ldots, K\omega\}, \text{~where} \quad \omega = \frac{2\pi}{K},
\end{equation}
with $K$ being the number of phase quantization levels. Consequently, the achievable SNR at the Rx is given by:
\begin{equation}
\begin{split}
\mathrm{SNR} &= \frac{P \left| \beta_{0}e^{j\alpha_0} + \sum_{n=1}^{N} \beta_{n} e^{j(\alpha_n+\theta_{n})} \right|^{2}}{\sigma^{2}}.
\end{split}
\label{eq:snr}
\end{equation}

Acquiring accurate CSI for IRS-aided systems is challenging, as it typically requires access to in-phase/quadrature (I/Q) signal components, which is not supported by current network protocols \cite{CSM_MIMO}. Therefore, we develop a blind IRS beamforming approach that relies solely on received signal power measurements, avoiding the need for explicit CSI estimation.
To collect the necessary data, the Rx configures the IRS with random phase-shift vectors over $T$ independent trials. Let $\boldsymbol{\theta}_t = \{\theta_{1t}, \theta_{2t}, \dots, \theta_{Nt}\}$ denote the phase configuration in the $t$-th trial\footnote{In practice, to improve sampling efficiency, $\boldsymbol{\theta}_t$ can be selected from a predefined codebook rather than generated randomly on the fly.}. The corresponding received signal power, $|Y_t|^2 = Y_t Y_t^{\mathrm{H}}$, is measured at the Rx. Thus, the available dataset consists of $T$ samples: $\left\{ \left[ \boldsymbol{\theta}_t, |Y_t|^2 \right] \right\}$ for $t = 1, 2, \dots, T$. The objective of the proposed blind IRS beamforming scheme is to determine the IRS phase-shift vector $\boldsymbol{\theta} = [\theta_1, \theta_2, \dots, \theta_N]$ that maximizes the SNR in (\ref{eq:snr}), using only this sampled dataset.

\section{Existing Schemes}
\label{sec:benchmark_schemes}
Before detailing the proposed blind beamforming method, we first introduce three existing schemes which serve as benchmarks. These schemes represent different approaches to IRS configuration, relying on varying levels of channel knowledge.

\subsection{CPP with CSI}
\label{subsec:cpp_with_csi}
Given perfect CSI, i.e., $h_n = \beta_n e^{j\alpha_n}$ for $n \in \mathcal{N} \cup \{0\}$, the CPP method provides a performance upper bound. It first finds the optimal solution in the ideal case of continuous phase shifts. For each IRS element $\mathrm{IRS}_n$, the optimal continuous phase $\theta_n^*$ is the one that aligns its reflected channel $h_n e^{j\theta_n}$ with the background channel $h_0$:
$$
\theta_n^* = \Delta_n \triangleq \alpha_0 - \alpha_n.
$$
Given the practical hardware constraint of discrete phase shifts from the set $\Phi_K$ defined in \eqref{eq:phase_set}, CPP projects this ideal solution to the nearest feasible point:
$$
\theta_n^{\mathrm{CPP}} = \arg\min_{\theta_n \in \Phi_K} \left| \theta_n - \Delta_n \right|, \quad \forall n \in \mathcal{N}.
$$
The final IRS configuration is $\boldsymbol{\theta}^{\mathrm{CPP}} = [\theta_1^{\mathrm{CPP}}, \theta_2^{\mathrm{CPP}}, \dots, \theta_N^{\mathrm{CPP}}]$. While offering optimal performance, CPP's reliance on perfect, instantaneous CSI is often impractical in real systems.

\subsection{Random-Max Sampling (RMS)}
\label{subsec:random_max_sampling}
In stark contrast, the Random-Max Sampling (RMS) algorithm operates without any CSI. It relies solely on the dataset $\{[\boldsymbol{\theta}_t, |Y_t| ^2]\}_{t=1}^{T}$ collected from $T$ random configuration trials. After sampling, RMS selects the single phase-shift vector that yielded the highest received power:
$$
\boldsymbol{\theta}^{\mathrm{RMS}} = \boldsymbol{\theta}_{t^\star}, \quad \text{where } t^\star = \arg\max_{1 \leq t \leq T} |Y_t|^2.
$$
Thus, RMS is a simple heuristic that chooses the best-observed sample. Its performance heavily depends on the number of trials $T$. A larger $T$ increases the chance of sampling a near-optimal configuration at the cost of higher sampling overhead.

\subsection{CSM Algorithm}
\label{subsec:csm_algorithm}
The CSM algorithm \cite{CSM_2023} offers a more refined, data-driven approach. Instead of using just one sample like RMS, it leverages conditional statistics from the entire dataset to configure each IRS element independently. The procedure is as follows:
\begin{enumerate}
    \item For each IRS element $\mathrm{IRS}_n$ and each candidate discrete phase $\varphi \in \Phi_K$, compute the conditional sample mean of the received power:
    $$
    \widehat{E}\left[|Y|^2 \mid \theta_n = \varphi\right] = \frac{1}{ |\mathcal{T}_{n,\varphi}| } \sum_{t \in \mathcal{T}_{n,\varphi}} |Y_t|^2,
    $$
    where $\mathcal{T}_{n,\varphi} = \{ t : \theta_{nt} = \varphi \}$ is the set of trials where the phase of $\mathrm{IRS}_n$ was $\varphi$.
    
    \item For each $\mathrm{IRS}_n$ select the phase that yields the highest conditional sample mean:
    $$
    \theta_n^{\mathrm{CSM}} = \arg\max_{\varphi \in \Phi_K} \widehat{E}\left[|Y|^2 \mid \theta_n = \varphi\right].
    $$
    The final configuration is $\boldsymbol{\theta}^{\mathrm{CSM}} = [\theta_1^{\mathrm{CSM}}, \theta_2^{\mathrm{CSM}}, \dots, \theta_N^{\mathrm{CSM}}]$.
\end{enumerate}
Theoretically, with a sufficient number of samples ($T = \Omega(N^{2}(\log N)^{3})$), the CSM configuration converges with high probability to the CPP solution \cite{CSM_2023}. However, in practice with limited $T$, the algorithm's core operation, comparing the magnitudes of the conditional means as in Step 2, only partially utilizes the information in the sampled data. This can limit its performance, especially when the number of phase quantization levels $K$ is large.

\section{Proposed Blind Beamforming Scheme}
\label{sec:proposed_scheme}
This section details the proposed blind beamforming scheme, termed Phase Estimation (PE) based beamforming. The core concept is to directly estimate the relative phase difference between the background channel and each IRS-reflected channel using only sampled received signal power measurements. Once estimated, the IRS phase shifts are configured to compensate for these differences, enabling coherent signal combining at the receiver. The scheme operates exclusively on the sampled dataset $\{[\boldsymbol{\theta}_t, |Y_t|^2]\}_{t=1}^{T}$, requiring no prior CSI.

\subsection{Problem Statement and Theoretical Foundation}
From the system model in Section \ref{sec:system_model}, the unconditional expectation of the received signal power, with each $\theta_n$ uniformly and independently drawn from $\Phi_K$, is given by:
\begin{equation}
\mathbb{E}[|Y|^{2}] = \beta_{0}^{2}P + \sum_{m=1}^{N}\beta_{m}^{2}P + \sigma^{2}.
\end{equation}
Conditioned on a specific phase shift $\theta_n = k\omega$ for the $n$-th IRS element, the conditional expectation is:
\begin{equation}
\mathbb{E}[|Y|^{2}|\theta_{n}=k\omega]=P|h_{0}+h_{n}e^{jk\omega}|^{2}+\sum_{m\neq n}\beta_{m}^{2}P+\sigma^{2}.
\end{equation}
Define the difference between these two expectations as:
\begin{equation}
J_{nk} \triangleq \mathbb{E}[|Y| ^{2} \mid \theta_n = k\omega] - \mathbb{E}[ |Y|
^{2}].
\end{equation}
Substituting the expressions and simplifying yields the key relationship:
\begin{equation}
J_{nk} = 2\beta_0\beta_n P \cos(k\omega - \Delta_n),
\label{eq:J_theoretical}
\end{equation}
where $\Delta_n = \alpha_0 - \alpha_n$ is the phase difference between the background channel and the $n$-th equivalent cascaded channel. The term $2\beta_0\beta_n P$ is an unknown positive amplitude. Crucially, $J_{nk}$ depends on the unknown $\Delta_n$ through a cosine function. The objective is to estimate $\Delta_n$ from noisy empirical observations $\{\widehat{J}_{nk}\}_{k=1}^{K}$.

\subsection{From Sampled Data to Phase Difference Estimation}
Using the sampled dataset, we first compute the empirical unconditional and conditional sample means:
\begin{align}
\widehat{E}[|Y| ^2] &= \frac{1}{T} \sum_{t=1}^{T} |Y_t|
^2, \\
\widehat{E}[|Y| ^2 \mid \theta_n = k\omega] &= \frac{1}{ |\mathcal{T}_{n,k}| } \sum_{t \in \mathcal{T}_{n,k}} |Y_t|
^2,
\end{align}
where $\mathcal{T}_{n,k} = \{t: \theta_{nt} = k\omega\}$. The empirical observation is then:
\begin{equation}
\widehat{J}_{nk} = \widehat{E}[|Y| ^2 \mid \theta_n = k\omega] - \widehat{E}[ |Y|
^2].
\label{eq:J_empirical}
\end{equation}
The relationship between the observation and the parameter $\Delta_n$ can be modeled as $\widehat{J}_{nk} = 2\beta_0\beta_n P \cos(k\omega - \Delta_n) + \epsilon_k$, where $\epsilon_k$ represents observation noise. Expanding the cosine term yields a linear model:
\begin{equation}
\begin{split}
\widehat{J}_{nk} =& \underbrace{(2\beta_0\beta_n P\cos\Delta_n)}_{A} \cos(k\omega) \\
&+ \underbrace{(2\beta_0\beta_n P\sin\Delta_n)}_{B} \sin(k\omega) + \epsilon_k.    
\end{split}
\label{eq:linear_model}
\end{equation}
Let $A$ and $B$ denote the two unknown intermediate coefficients. A linear least-squares problem is formulated to estimate $A$ and $B$ from the $K$ observations $\{\widehat{J}_{nk}\}_{k=1}^{K}$:
\begin{equation}
\min_{A, B} \sum_{k=1}^{K} \left[ A \cos(k\omega) + B \sin(k\omega) - \widehat{J}_{nk} \right]^2.
\label{eq:LS_obj}
\end{equation}
Exploiting the orthogonality of the cosine and sine bases for the standard discrete phase set $\Phi_K$ (where $\omega=2\pi/K$), i.e.,
\begin{equation}
\begin{split}
\sum_{k=1}^{K} \cos^2(k\omega) = \sum_{k=1}^{K} \sin^2(k\omega) = \frac{K}{2}, \\ \quad \sum_{k=1}^{K} \cos(k\omega)\sin(k\omega) = 0,    
\end{split}
\end{equation}
the least-squares solutions are obtained in closed form:
\begin{align}
A^* &= \frac{2}{K} \sum_{k=1}^{K} \widehat{J}_{nk} \cos(k\omega), \label{eq:A_solution} \\
B^* &= \frac{2}{K} \sum_{k=1}^{K} \widehat{J}_{nk} \sin(k\omega). \label{eq:B_solution}
\end{align}
From the definitions in \eqref{eq:linear_model}, the phase difference $\Delta_n$ satisfies $\tan\Delta_n = B/A$. Therefore, an estimate $\hat{\Delta}_n$ is obtained using the four-quadrant arctangent function\footnote{The amplitude of the reflection channel is estimated as $\hat{\beta}_n = \frac{\sqrt{(A^*)^2 + (B^*)^2}}{2 \beta_0 P}$, assuming $\beta_0 P$ is known a priori. These estimates can be leveraged in broader communication scenarios, which are reserved for our future study.}:
\begin{equation}
\hat{\Delta}_n = \operatorname{atan2}(B^*, A^*).
\label{eq:Delta_estimate}
\end{equation}
This estimate requires only the empirical observations $\{\widehat{J}_{nk}\}$.

\subsection{IRS Phase Configuration Based on Phase Estimation}
The optimal continuous phase shift for the $n$-th IRS element to align its reflected signal with the background channel is $\theta_n^* = \Delta_n$. Given the estimate $\hat{\Delta}_n$ and the discrete phase constraint $\theta_n \in \Phi_K = \{\omega, 2\omega, \dots, K\omega\}$, we select the discrete phase that most closely compensates for the estimated phase difference. The procedure is as follows:
\begin{enumerate}
    \item Map the estimated phase difference $\hat{\Delta}_n \in (-\pi, \pi]$ to the range $[0, 2\pi)$: $\hat{\Delta}_n' = \mathrm{mod}(\hat{\Delta}_n, 2\pi)$.
    \item For each discrete phase value \(k\omega\) (\(k = 1, \dots, K\)), compute the circular distance $d_k = \min\!\bigl(|\hat{\Delta}_n' - k\omega|,\; 2\pi - |\hat{\Delta}_n' - k\omega|\bigr)$, and find the index that minimizes it: $k_n^* = \arg\min_{k \in \{1,\dots,K\}} d_k$.
    \item Set the phase shift for the $n$-th element to $\theta_n^{\mathrm{PE}} = k_n^* \omega$.
\end{enumerate}
Applying this process to all $N$ IRS elements yields the final phase configuration vector:
\begin{equation}
\boldsymbol{\theta}^{\mathrm{PE}} = [\theta_1^{\mathrm{PE}}, \theta_2^{\mathrm{PE}}, \dots, \theta_N^{\mathrm{PE}}].
\label{eq:PE_config}
\end{equation}
This configuration $\boldsymbol{\theta}^{\mathrm{PE}}$ aims to co-phase the signals via the direct and IRS-reflected paths, thereby maximizing the received signal power and the SNR in \eqref{eq:snr}. The complete PE-based blind beamforming algorithm is summarized in Algorithm \ref{alg:PE}.

\newtheorem{remark}{Remark}
\begin{remark}[Equivalence for $K=2$]
\label{Equivalence for K=2}
\rm{
When the number of phase quantization levels $K=2$, the proposed algorithm is theoretically equivalent to the CSM algorithm. Specifically, from \eqref{eq:linear_model}, \eqref{eq:A_solution} and \eqref{eq:B_solution}, we have $\widehat{J}_{n1} = -(2\beta_0\beta_n P\cos\Delta_n +\epsilon_1)$ and $\widehat{J}_{n2} = 2\beta_0\beta_n P\cos\Delta_n +\epsilon_2$. Consequently, $A^* = \widehat{J}_{n2} - \widehat{J}_{n1} = 4\beta_0\beta_n P\cos\Delta_n + (\epsilon_2-\epsilon_1)$ and $B^* = 0$. Thus, the estimation formula \eqref{eq:Delta_estimate} reduces to:
\begin{equation}
\label{equivalent}
                \hat {\Delta} _n=\left\{\begin{array}{cc}
                2\pi~ \text{or} ~0, & \text{if~} \widehat{J}_{n2}>\widehat{J}_{n1} \\
                \pi, & \text{otherwise}
                \end{array}\right.,  
\end{equation}
which is equivalent to the decision rule of the CSM algorithm.
}
\end{remark}

\begin{remark}[Computational Complexity]
\label{Computational Complexity}
\rm{
The initial steps of Algorithm \ref{alg:PE} (computing the conditional sample means) are identical to those of the CSM algorithm \cite{CSM_2023}. The key difference lies in the subsequent steps (8-10), where the proposed method constructs a least-squares estimate for $\Delta_n$. Both algorithms share the same computational complexity of $\mathcal{O}(TNK)$, as the additional operations for solving the least-squares problem are only $\mathcal{O}(K)$ per element.
}
\end{remark}

\begin{algorithm}[t]
\caption{Proposed IRS Blind Beamforming}
\label{alg:PE}
\begin{algorithmic}[1]
\REQUIRE Received power samples $\{[\boldsymbol{\theta}_t, |Y_t|^2]\}_{t=1}^{T}$, number of IRS elements $N$, number of phase levels $K, \omega = 2\pi/K$.
\ENSURE IRS phase configuration $\boldsymbol{\theta}^{\mathrm{PE}}$.
\STATE Compute the unconditional sample mean:\\
$\widehat{E}[|Y| ^2] = \frac{1}{T}\sum_{t=1}^{T} |Y_t|^2$.
\FOR{each IRS element $n$ = 1 to $N$}
    \FOR{each phase level $k$ = 1 to $K$}
        \STATE Identify sample set $\mathcal{T}_{n,k} = \{t: \theta_{nt} = k\omega\}$.
        \STATE Compute conditional sample mean:\\ $\widehat{E}[|Y| ^2 \mid \theta_n = k\omega] = \frac{1}{ |\mathcal{T}_{n,k}| }\sum_{t \in \mathcal{T}_{n,k}} |Y_t|^2$.
        \STATE Compute $\widehat{J}_{nk} = \widehat{E}[|Y| ^2 \mid \theta_n = k\omega] - \widehat{E}[ |Y|^2]$.
    \ENDFOR
    \STATE Compute $A^* = \frac{2}{K} \sum_{k=1}^{K} \widehat{J}_{nk} \cos(k\omega)$.
    \STATE Compute $B^* = \frac{2}{K} \sum_{k=1}^{K} \widehat{J}_{nk} \sin(k\omega)$.
    \STATE Estimate phase difference: $\hat{\Delta}_n = \operatorname{atan2}(B^*, A^*)$.
    \STATE Map to $[0, 2\pi)$: $\hat{\Delta}_n' = \mod(\hat{\Delta}_n, 2\pi)$.
    \STATE Find nearest discrete phase: \\ $k_n^* = \arg\min_{k \in \{1,\dots,K\}} \left| \hat{\Delta}_n' - k\omega \right|$.
    \STATE Set $\theta_n^{\mathrm{PE}} = k_n^* \omega$.
\ENDFOR
\STATE \textbf{return} $\boldsymbol{\theta}^{\mathrm{PE}} = [\theta_1^{\mathrm{PE}}, \theta_2^{\mathrm{PE}}, \dots, \theta_N^{\mathrm{PE}}]$.
\end{algorithmic}
\end{algorithm}

\section{Numerical Results}
\label{sec:simulations}
This section evaluates the achievable SNR of the proposed IRS blind beamforming scheme against several benchmark methods, focusing on the impacts of three key system parameters: $K$, $N$, and $T$.

\subsection{Simulation Setup}
\label{subsec:sim_setup}
We consider a carrier frequency of $f = 3$ GHz. The locations of Tx, IRS, and Rx are fixed at Cartesian coordinates [50, -100, 20], [-10, -10, 0], and [0, 0, 0] meters, respectively. Unless otherwise specified, the default system parameters are set as: $P = 30$ dBm, $\sigma^2 = -90$ dBm, $N = 200$, $K = 8$, and $T = 2000$. The background channel $h_0$ and the IRS-cascaded channels $h_n$ are modeled as described below.

\textbf{Antenna and IRS Gains:}
\begin{itemize}
    \item \textbf{Tx and Rx:} Both are equipped with isotropic antennas, yielding gains $G_t = 1$ and $G_r = 1$.
    \item \textbf{IRS Element:} Each passive element is modeled as a square patch with a side length of $\lambda/2$, where $\lambda$ is the carrier wavelength.
    \begin{itemize}
        \item In the receiving link (Tx-to-IRS), its effective aperture is $A_{\text{irs,rx}} = (\lambda/2)^2$, corresponding to a receiving gain of $G_{\text{irs,rx}} = {4\pi A_{\text{irs,rx}}}/{\lambda^2} = \pi$.
        \item In the reflecting link (IRS-to-Rx), it is approximated as an isotropic radiator, yielding a gain of $G_{\text{irs,tx}} = 1$.
    \end{itemize}
\end{itemize}

\textbf{Large-Scale Fading (Path Loss):}
The large-scale path loss are calculated based on the distances and the respective antenna gains.
\begin{itemize}
    \item \textbf{Background Channel ($h_0$):} The large-scale fading amplitude is given by
    \begin{equation}
        \beta_0^{\text{(LS)}} = \frac{\lambda}{4\pi d_0^{a_0/2}} \sqrt{G_t G_r} = \frac{\lambda}{4\pi d_0^{a_0/2}},
    \end{equation}    
    where $d_0$ is the Tx-Rx distance, and $a_0$ is the associated path loss exponent.
    \item \textbf{Cascaded Channel ($h_n$):} The large-scale fading amplitude is the product of the amplitudes for the two hops. The first hop (Tx to IRS element) amplitude is
    \begin{equation}
    \beta^{\text{(LS1)}} = \frac{\lambda}{4\pi d_1^{a_1/2}} \sqrt{G_t \cdot G_{\text{irs,rx}}} = \frac{\lambda\sqrt{\pi}}{4\pi d_1^{a_1/2}},
    \end{equation}
    where $d_1$ is the Tx-IRS distance, and $a_1$ is its path loss exponent. The second hop (IRS element to Rx) amplitude is
    \begin{equation}
    \beta^{\text{(LS2)}} = \frac{\lambda}{4\pi d_2^{a_2/2}} \sqrt{G_{\text{irs,tx}} \cdot G_r} = \frac{\lambda}{4\pi d_2^{a_2/2}},
    \end{equation}
    where $d_2$ is the IRS-Rx distance, and $a_2$ is its path loss exponent. Consequently, the overall cascaded large-scale fading amplitude is
    \begin{equation}
    \beta_n^{\text{(LS)}} = \beta^{\text{(LS1)}} \cdot \beta^{\text{(LS2)}} = \frac{\lambda^2 \sqrt{\pi}}{(4\pi)^2 d_1^{a_1/2} d_2^{a_2/2}}.
    \end{equation}
\end{itemize}
In our simulations, the Rx is situated in a weak coverage area of the Tx, a typical deployment scenario for an IRS. Therefore, the background link is modeled with a higher path loss exponent, $a_0 = 4$. By strategically deploying the IRS, a good channel condition is assumed for the cascaded link. Hence, we set $a_1 = a_2 = 2$ for the Tx-IRS and IRS-Rx links. The distances $d_0$, $d_1$, and $d_2$ are calculated from the fixed node coordinates provided above.

\textbf{Small-Scale Fading:}
The small-scale fading components are modeled as independent and identically distributed (i.i.d.) unit complex Gaussian random variables, i.e., $\mathcal{CN}(0,1)$.

\subsection{Benchmark Schemes}
The performance of the proposed algorithm is compared against the following benchmark schemes:
\begin{itemize}
    \item \textbf{CPP with Perfect CSI:} This scheme, described in Section~\ref{subsec:cpp_with_csi}, assumes perfect knowledge of all channel coefficients $\{h_n\}_{n=0}^N$. It computes the ideal continuous phase shifts $\theta_n = \alpha_0 - \alpha_n$ and projects them to the nearest points in the discrete set $\Phi_K$. This serves as the performance upper bound.
    \item \textbf{CSM:} This benchmark, detailed in Section~\ref{subsec:csm_algorithm}, selects for each IRS element the discrete phase shift that yields the highest conditional sample mean of the received power, computed from the sampled dataset.
    \item \textbf{RMS:} As described in Section~\ref{subsec:random_max_sampling}, this method selects the single IRS phase-shift vector (from the $T$ trials) that corresponds to the highest instantaneous received power measurement.
    \item \textbf{Random Phase Configuration:} Each IRS element is configured with a phase shift uniformly and independently selected from the discrete set $\Phi_K$. This represents a naive, non-adaptive strategy.
    \item \textbf{Without IRS:} The baseline SNR achieved via the direct Tx-Rx link $h_0$ only, without the IRS.
\end{itemize}
The average SNR for each scheme is evaluated over 1000 independent Monte Carlo trials.


\subsection{Performance Evaluation and Discussion}
Based on the simulation setup described above, the performance of the proposed and benchmark schemes is evaluated with the following discussions.

\begin{figure}[t]
\centering
\includegraphics[width=0.5\textwidth]{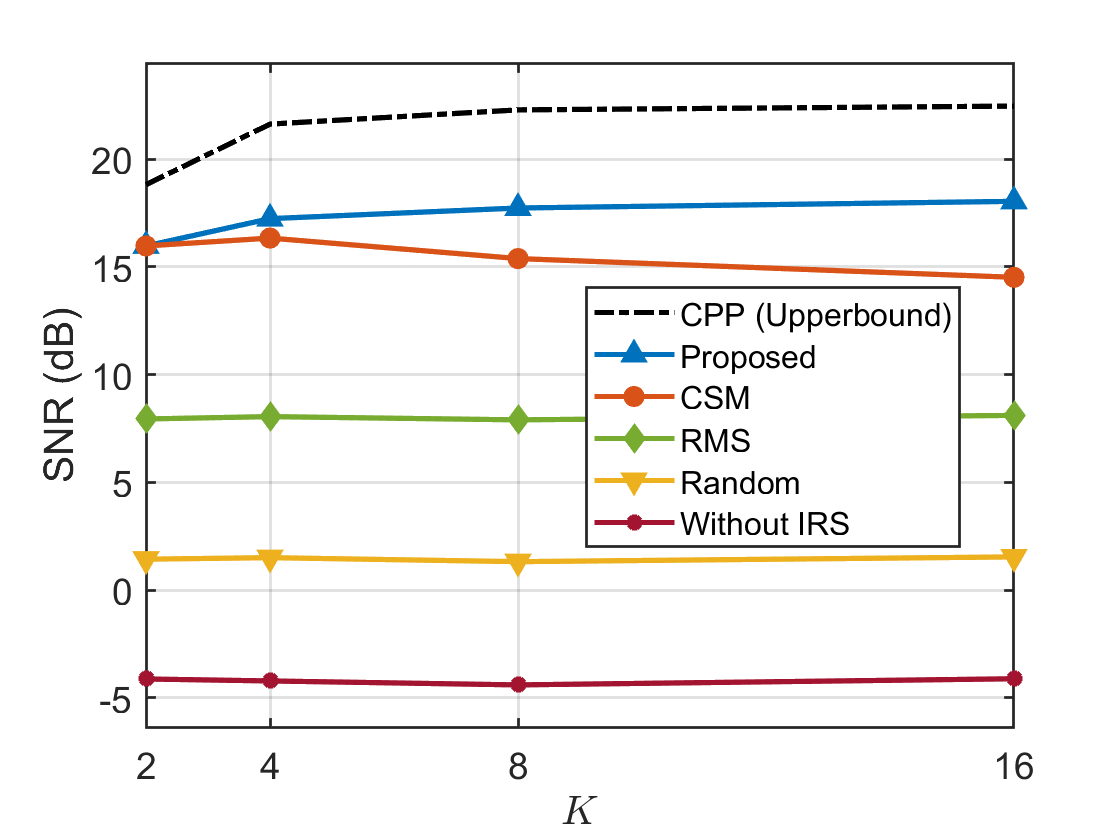}
\caption{$\mathrm{SNR}$ vs. $K$.}
\label{fig_SNR_K}
\end{figure}
\begin{figure}[t]
\centering
\includegraphics[width=0.5\textwidth]{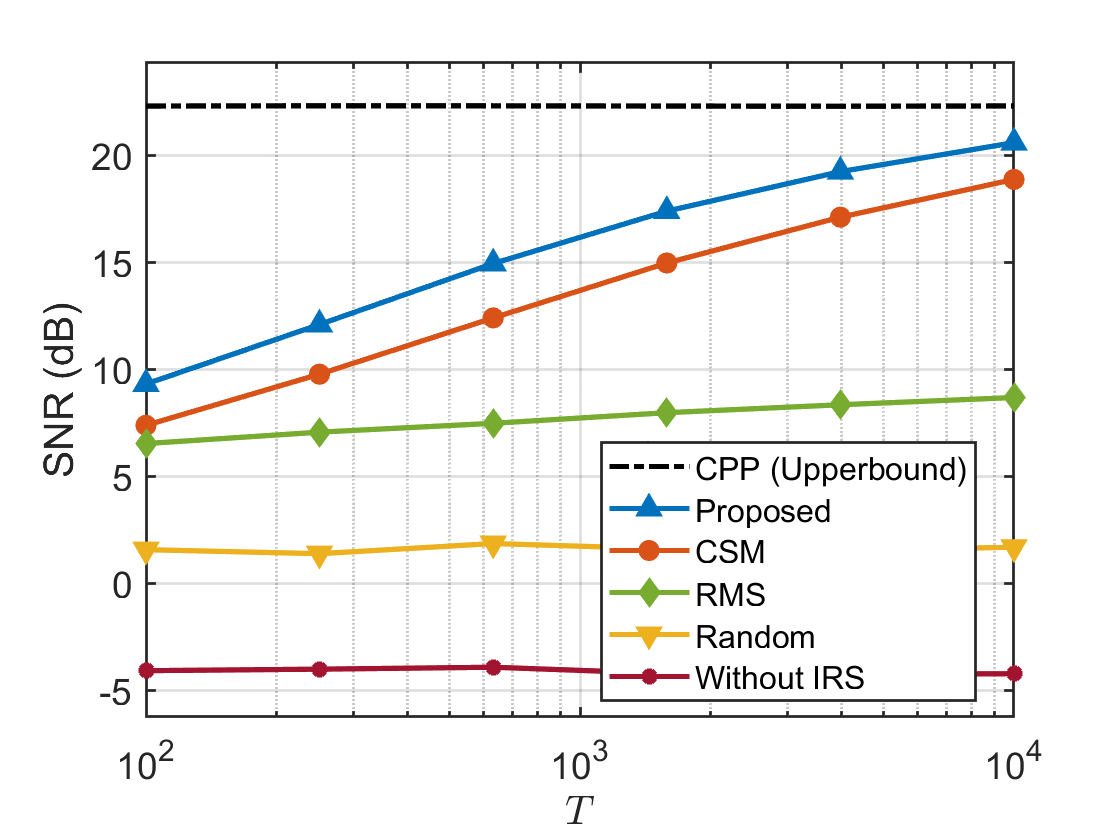}
\caption{$\mathrm{SNR}$ vs. $T$.}
\label{fig_SNR_T}
\end{figure}

Fig.~\ref{fig_SNR_K} illustrates the average SNR performance as a function of the number of phase quantization levels $K$. A key observation is that when $K=2$, the performance of the proposed algorithm is identical to that of the CSM algorithm. This validates the theoretical equivalence derived in Remark \ref{Equivalence for K=2}. As $K$ increases, the performance gain of the proposed algorithm over CSM becomes more pronounced, e.g., $+ 2.4$ dB and $+ 3.5$ dB for $K = 8$ and 16, respectively. This is because the CSM algorithm only partially utilizes the sample information by selecting the phase corresponding to the maximum value from the set $\{\widehat{J}_{nk}\}_{k=1}^{K}$. In contrast, the proposed algorithm fully exploits the information within the sampled data by solving a least squares problem to estimate the relative phase difference $\Delta_n$, leading to a more accurate phase configuration. Furthermore, the performance gap between the proposed algorithm and the CPP upper bound is attributable to the limited number of samples $T$; this relationship is explored further in Fig.~\ref{fig_SNR_T}. Finally, as expected, the RMS, Random, and No-IRS schemes show no sensitivity to variations in $K$.

\begin{figure}[t]
\centering
\includegraphics[width=0.5\textwidth]{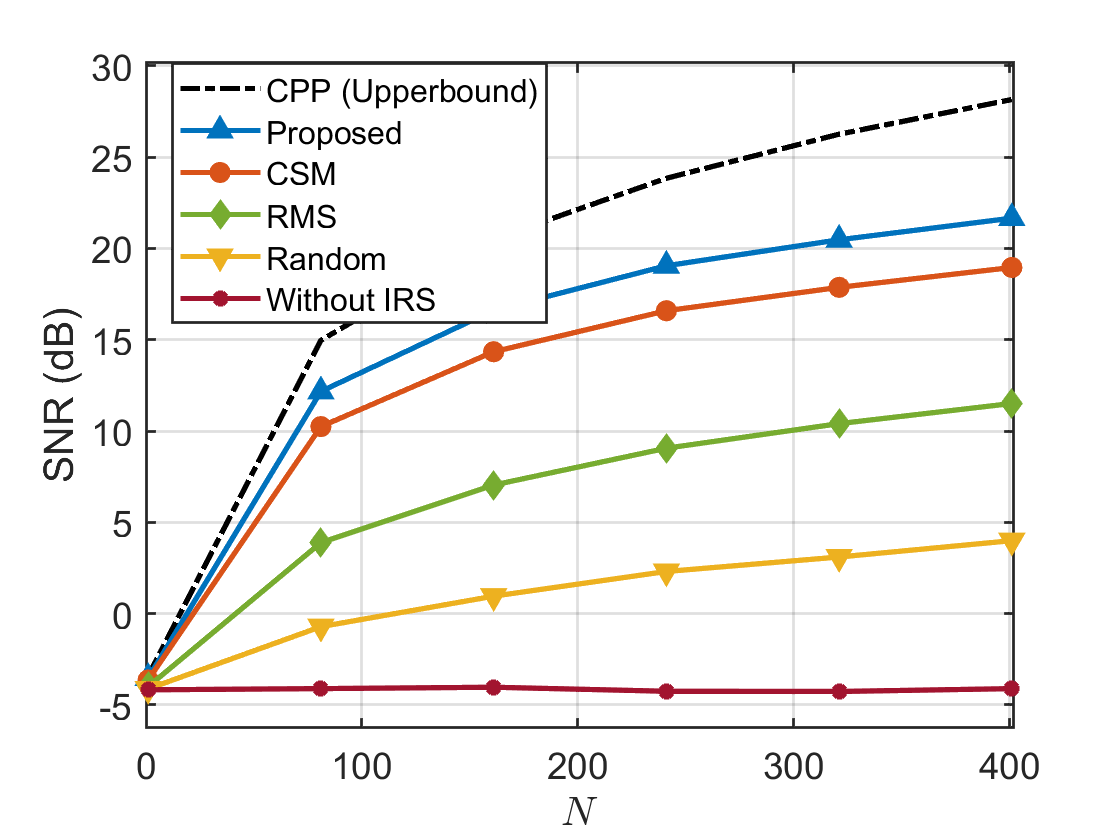}
\caption{$\mathrm{SNR}$ vs. $N$.}
\label{fig_SNR_N}
\end{figure}
Fig.~\ref{fig_SNR_N} depicts the SNR performance versus the number of IRS elements $N$. The proposed scheme consistently outperforms all other CSI-free benchmarks (CSM, RMS, Random) across different values of $N$. However, the performance gap between the proposed scheme and the CPP upper bound widens as $N$ increases. This trend occurs because as $N$ increases, the fixed number of samples $T$ increasingly limits the performance of the proposed method.

Fig.~\ref{fig_SNR_T} shows the impact of the number of sampling trials $T$ on the achievable SNR. The proposed scheme maintains a consistent performance advantage over other benchmark schemes for all evaluated $T$. Crucially, as $T$ increases, the performance of the proposed method converges towards the CPP upper bound, empirically validating its asymptotic effectiveness. This confirms that with sufficient samples, the proposed algorithm can accurately estimate the differences $\Delta_n$. In contrast, while the performance of the RMS method also improves with larger $T$, its rate of improvement is significantly slower. This is due to RMS's inefficient, brute-force sampling strategy, which fails to systematically extract and leverage the channel information embedded in the data as the proposed least-squares-based approach does.

\section{Conclusion}
\label{sec:conclusion}
We proposed a novel blind IRS beamforming algorithm that estimates the phase differences between channels by solving a least-squares problem, using only received signal power samples. 
The proposed scheme is proven equivalent to CSM for binary phase shifts ($K=2$) but delivers superior performance for $K > 2$, as verified by simulations. The performance gain increases with $K$, while computational complexity remains at the same order, $\mathcal{O}(TNK)$. Future work may extend the approach to multi-user and broadband scenarios.

\def\baselinestretch{1}
\bibliographystyle{IEEEbib}
\bibliography{IEEErefs}
\end{document}